# Automated charting of the visual space of insect compound eyes

Mauricio Muñoz Arias, John K. Douglass, Martin F. Wehling and Doekele G. Stavenga

*Abstract—* This paper describes the automatic measurement of the spatial organization of the visual axes of insect compound eyes, which consist of several thousands of visual units called ommatidia. Each ommatidium samples the optical information from a small solid angle, with an approximately Gaussian-distributed sensitivity (half-width of the order of 1°) centered around a visual axis. Together, the ommatidia gather the optical information from virtually the full surroundings. The spatial distribution of the visual axes thus determines the eye's spatial resolution. Knowledge of the optical organization of a compound eye and its visual acuity is crucial for quantitative studies of the neural processing of visual information. Here we present an automated procedure for mapping a compound eye's visual axes, using an intrinsic optical phenomenon, the pseudopupil. We outline the optomechanical setup for scanning insect eyes, and use experimental results obtained from a housefly, *Musca domestica*, to illustrate the steps in the measurement procedure.

**Note to practitioners** - Mapping of the visual axis distribution of compound eyes has previously been performed manually, involving extremely tedious procedures. The development of an automated analysis of the optical organization of compound eyes allows assembling the necessary data in a much more effective, time-saving way. The necessary experimental steps require a number of algorithms that we have developed using Matlab. The optical and mechanical elements of the experimental setup are mostly standard.

## I. Introduction

INSECT compound eyes have been recognized as powerful optical devices enabling acute and versatile visual capacities [1, 2]. The compactness of insect visual systems and the agility of their owners, demonstrating highly developed processing of optical information, has intrigued numerous researchers as well as laymen. Flies, for instance, are well-known for their fast responses to moving objects, and bees are famous for possessing color vision as well as polarization vision [2].

The compound eyes of arthropods consist of numerous, anatomically similar units, the ommatidia, each of which is capped by a facet lens. The assembly of facet lenses, known collectively as the cornea, usually approximates a hemisphere. Each ommatidium samples incident light from a small solid angle with half-width of the order of 1°. The ommatidia of the two eyes together sample about the full solid angle, but the visual axes of the ommatidia are not evenly distributed. Certain eye areas have a high density of visual axes, which creates an area of high spatial acuity, colloquially called a fovea. The remaining part of the eye then has a coarser spatial resolution [3-9].

Knowledge of the optical organization of the compound eyes is crucial for quantitative studies devoted to unraveling the neural processing of visual information. Studies of the neural networks of an insect's brain (e.g. [10]) therefore often require knowledge of the spatial distribution of the ommatidial axes. Furthermore, compound eyes have inspired several technical innovations. Many initiatives to produce bioinspired artificial eyes have built on existing quantitative studies of real compound eyes [11-13]. For instance, a semiconductor-based sensor with high-spatial resolution was designed based on the model of insect compound eyes [11, 14-17]. However, the devices developed so far have not implemented the actual characteristics of existing insect eyes. Accurate representations of insect compound eyes and their spatial organization will require solid data from real eyes, which is not extensively available.

The main reason for the paucity of data is the extreme tediousness of the available procedures for charting the eyes' spatial characteristics. This motivated attempts to establish a more automated eye mapping procedure. In a first attempt at automated analyses of insect compound eyes, Douglass and Wehling [18] developed a scanning procedure for mapping facet sizes in the cornea and demonstrated its feasibility for a few fly species. Here we extend their approach by developing methods for not only scanning the facets of the cornea, but also assessing the visual axes of the ommatidia to which the facets belong. We present the case of housefly eyes to exemplify the procedures involved.

This study was financially supported by the Air Force Office of Scientific Research/European Office of Aerospace Research and Development AFOSR/EOARD (grant FA9550-15-1-0068, to D.G.S.). We thank Kehan Satu, Hein Leertouwer, and Oscar Rincón Cardeño for assistance.

Mauricio Muñoz Arias (corresponding author), is with the Design Group, Faculty of Science and Engineering, University of Groningen, Nijenborgh 4, NL9747AG Groningen, The Netherlands (email: m.munoz.arias@rug.nl).

John K. Douglass and Martin F. Wehling are with the Air Force Research Laboratory, Eglin Air Force Base, 101 W. Eglin Blvd., FL 32542, USA (emails: jdcentro@gmail.com and fribbit@gmail.com).

Doekele G. Stavenga is with the Surfaces and thin films department, Zernike Institute for Advanced Materials, University of Groningen, Nijenborgh 4, NL9747AG Groningen, The Netherlands (email: d.g.stavenga@rug.nl).

The experimental setup for scanning insect eyes is partly optical, i.e. a microscope with camera and illumination optics, partly mechanical, i.e. a goniometer system for rotating the investigated insect, and partly electronic, i.e. the drivers for the instruments and the computational programs for running the experiments. The developed methods encompass a range of computational procedures, from capturing images, choosing camera channels, and setting image processing thresholds, to recognizing individual facet locations via bright spots of light reflected from their convex surfaces. Fourier transform methods were crucial in the image analysis, both for detecting individual facets, and for analyzing the facet patterns.

The paper is structured as follows. We first introduce the experimental setup and the optical marker, the pseudopupil, that is used to identify the visual axes of the eye's photoreceptors. Subsequently the algorithms that are used in the scanning procedure and image analysis are outlined.

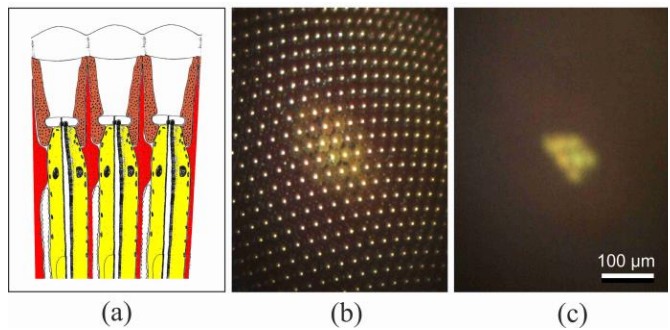

Fig. 1. Optics of fly eyes. (a) Diagram of three ommatidia of a fly eye, each capped by a biconvex facet lens, which focusses incident light onto a set of photoreceptor cells (yellow), surrounded by primary (brown) and secondary (red) pigment cells. Intense illumination causes migration of yellow pigment granules, which exist inside the photoreceptor cells, toward the tip of the photoreceptors, where they absorb and backscatter light. (b) Image at the level of the eye surface, showing the facet reflections (dots) as well as the pigment granule reflection in the activated state (the corneal pseudopupil, CPP). (c) Image taken at the level of the center of eye curvature (the deep pseudopupil, DPP). The photoreceptor cells are arranged in a seven-dot trapezoid, with their distal ends positioned at about the focal plane of the facet lenses. A superimposed virtual image of the photoreceptor tips thus exists in the plane of the center of eye curvature.

## II. MATERIALS AND METHODS

### A. Fly eyes and pseudopupils

Insect eyes consist of numerous more or less identical building blocks, called ommatidia, which consist of a facet lens and a set of photoreceptor cells enveloped by pigment cells. Incident light focused at the photoreceptors activates a pupil mechanism, a system of mobile, yellow-colored pigment granules inside the photoreceptor cells, which controls the light flux that triggers the phototransduction process in the photoreceptors. This system thus has essentially the same function as the pupil in the human eye [19, 20]. The activation of the pupil mechanism causes a locally enhanced reflection in the eye area facing the aperture of the microscope's objective (Fig. 1). The position of the brightly reflecting eye area, called the pseudopupil [19-21], changes upon rotation of the eye, because then the incident light activates the pupil mechanism in a different set of photoreceptor cells. The pseudopupil thus acts as a marker of the ommatidia aligned with the microscope, which allows mapping of the spatial distribution of the eye's visual axes [20-24].

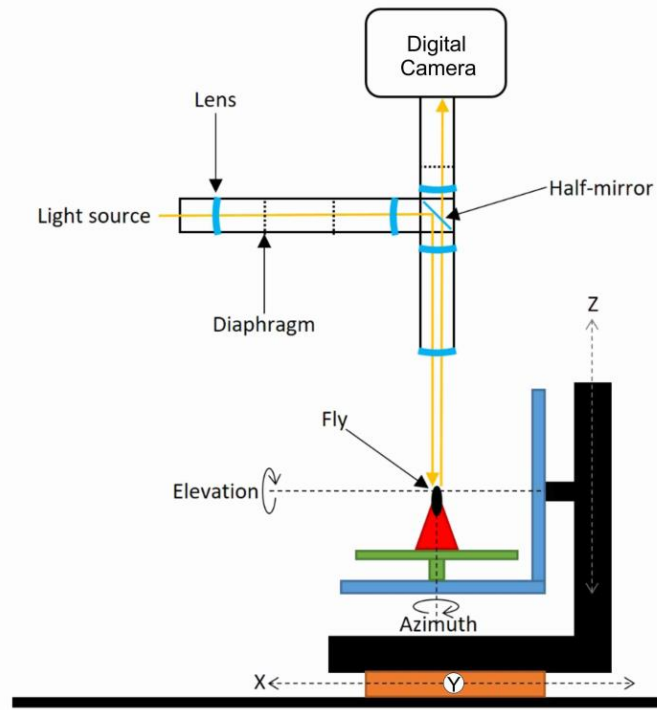

Fig. 2. GRACE, the Goniometric Research Apparatus for Compound Eyes. The investigated insect (a fly) is mounted at the motorized stage consisting of three translation stages (X, Y, Z) and two rotation stages (Elevation and Azimuth). Light from a light source is focused by a lens at a diaphragm, which is focused by a telescope via a half-mirror at the fly's eye. The eye is photographed with a camera attached to a microscope.

### B. Animals and optical stimulation

Experiments were performed on houseflies (*Musca domestica*) obtained from a culture maintained by the department of Evolutionary Genetics at the University of Groningen. Before the measurements, a fly was immobilized by gluing it with a low-melting point wax in a well-fitting tube. The fly was subsequently mounted on the stage of a motorized goniometer (Standa, Vilnius, Lithuania). The center of the two rotary stages coincided with the focal point of a telemicroscopic setup [25]. The epi-illumination light beam is supplied by a light source, which focuses light on a diaphragm that is focused at the fly's eye via a half-mirror in an optical telescope (Fig. 2). The optical axes of the ommatidia are assessed by rotating the fly in fixed steps and taking photographs after each step with a color digital camera attached to a microscope (Fig. 2). Because the pupillary pigment granules reflect predominantly in the long-wavelength range, the red channel of the digital camera is used to discriminate the pseudopupil from the facet lens reflections. The latter are best isolated from the pseudopupil by using the camera's blue channel (see Algorithms 1-3 in the Appendix).

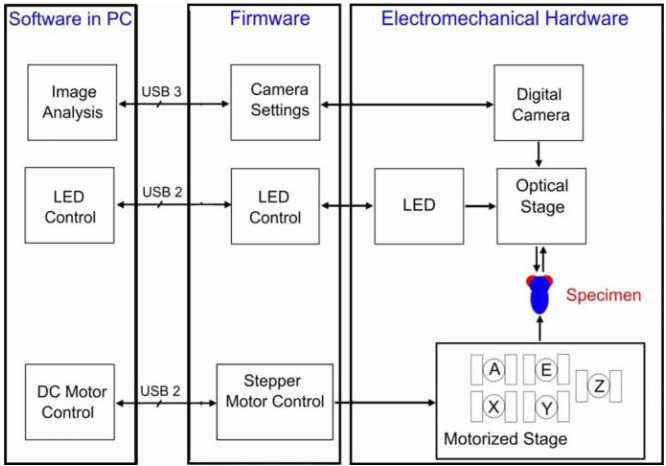

Fig. 3. Diagram of the GRACE system. The PC software controls the firmware, which drives the electromechanical hardware. The digital camera takes, via an optical stage, images of the specimen's eye, the LED light source illuminates the specimen, and the motors of the motorized stage actuate the X, Y-, and Z-translations and the azimuth (A) and elevation (E) rotations.

*C. The goniometric system*

In order to achieve alignment with the illumination, the fly's eye has to be photographed with the corneal facet lenses in focus, and the pseudopupil has to be centered. After each rotation, before photographing the fly has to be refocused and recentered. This is realized with the GRACE system (Goniometric Research Apparatus for Compound Eyes), which is diagrammatically shown in Fig. 3. It consists of three main subsystems: the lower and upper stages with their respective electronics as the hardware, the firmware embedded in the physical controllers, and the software with algorithms programmed in Matlab (PC). The hardware consists of the motorized and optical stages, the digital camera and a white LED light source. The firmware's routines are included in the Arduino controller for the stepper motors, for the intensity of the LED, and for the microcontroller of the digital camera (PointGrey, BFLY-U3-23S6C-C). The software consists of the algorithms for controlling motors and for capturing and analysing the acquired images. The information processed by the algorithms allows the control of the intensity of the LED and the position and speed of the stage's motors. The algorithms discussed next represent the major milestones that enable the GRACE system to scan insect eyes.

*D. Autofocusing and autocentering algorithms*

The main algorithms used when scanning an insect's eye are the autofocusing and autocentering algorithms. The goal of autofocusing is to bring the corneal level in focus of the camera, so to detect the facet reflections that are necessary for identification of the ommatidia (Fig. 1b). The procedure for detecting the corneal level is to change the vertical (Z-)position of the fly in steps and applying the Fast Fourier Transform (FFT) to the images taken at each level to determine the spatial frequency content of the images. The criterion for optimal focus then is the level with the greatest summed power above the low frequency cutoff.

The inputs for autofocusing (Algorithm 1) are camera streaming *vid* and Z-position; the outputs are the integral of the high-frequency content of the image $S_h$ and the focusing level $Z_h$ where $S_h$ is maximal. In the initial step, the Z-position of the camera image is adjusted to slightly below the corneal facet lenses, the image channel C is chosen and the region of interest (R) to determine the image's frequency content, i.e. $I_C$, is set. The *for* loop starts the image capture and the calculation of the sum of the high-pass filtered Fourier-transform $S_h$. By then stepping the Z-axis motor upwards, to an image level above the cornea, the level with the highest frequencies is found, i.e., where $S_h$ is maximal, which then is taken to be the corneal level $Z_h$. The Z-axis motor is then adjusted to the $Z_h$ level and an image is taken.

When focusing down from the cornea towards the level of the eye's center of curvature, the corneal facet reflections fade away and the pseudopupil reflections coalesce into a typical seven-dot pattern, which is characteristic for the organization of the photoreceptors within the fly ommatidia (Fig. 1c). The pattern at the level of the eye's center of curvature is called the deep pseudopupil [19]. In between the corneal and deep pseudopupil, the pupillary reflections create a more or less oval-shaped light spot. Shifting the fly with the X- and Y-motors so that the center of the light spot coincides with the center of the camera image is called autocentering. This procedure aligns the visual axis of the ommatidium whose facet is in the image center with the illumination beam and the optical axis of the microscope and camera system. In the autocentering procedure (Algorithm 2), the image is Gaussian filtered and binarized. The inputs are the camera streaming *vid* and the positions of the X- and Y-motors; the output is the distance between the centers of image and pseudopupil.

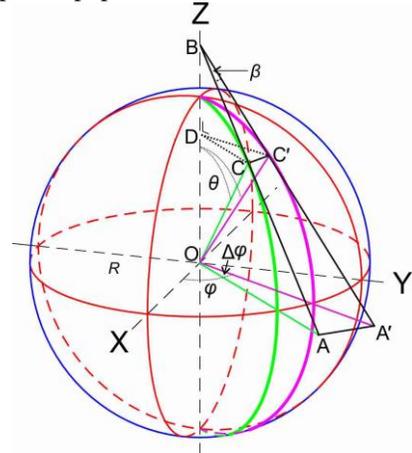

Fig. 4. Diagram for deriving the image rotation when scanning the fly eye.

*E. Correlating images*

The eye is scanned by taking and storing photographs at various values of the goniometer elevation $\theta$ and azimuth $\varphi$ after the autofocusing and autocentering procedures. To correlate the different images, it is important to realize that a change in $\varphi$ results in a rotation of the current image with respect to the previous image. For instance, let us assume that the center of an initial image corresponds to point C of a sphere (Fig. 4) and that a change in azimuth occurs so that plane OAB is rotated over a small angle $\Delta\varphi$, becoming plane OA'B. The center of the image then changes from point C to point C' (Fig. 4). If the X-axis of the camera image is perpendicular to the rotating plane, and thus the Y-axis is parallel to that plane, rotation of plane OAB to OA'B causes rotation of the image axes over an angle $\beta = \cos\theta\,\Delta\varphi$, as $\beta = CC'/BC$, $CC' = CD\Delta\varphi$, and $\cos\theta = CD/BC$ (Fig. 4). This means, that at the top ($\theta = 0°$), $\beta = \Delta\varphi$, and at the equator ($\theta = 90°$) $\beta = 0°$. When $\Delta\varphi = 0°$, that is, only the elevation $\theta$ is changed, the images are not rotated with respect to each other, or, $\beta = 0°$.

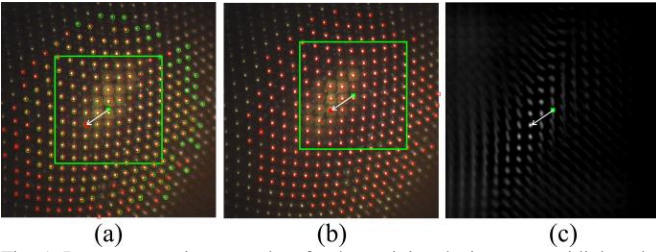

Fig. 5. Image processing procedure for determining the interommatidial angle. (a, b) Two successive images taken during a scan across the eye. (a) Image with facet centroids marked by green circles and a green dot at the image center. (b) Image after an azimuthal rotation of 5°, with facet centroids marked by red squares and a red dot at the image center. (c) Correlogram of the area within the green square of (a) correlated with image (b). The vector from the center of (c) to the maximum value of the correlogram represents the relative shift of images (a) and (b). Using that vector, the shifted square of (a) and its center are drawn in (b) and the facet centroids (red squares) of (b) are added in (a).

During the scanning procedure, after each rotation, the autocentering procedure centers the ommatidium with visual axis aligned with the optical axis of the measuring system, causing a rotation by an angle $\beta$ and a translation of the facet pattern. To determine the latter shift, we correlated two successive images (after first rotating the first image by $\beta$) as explained in Fig. 5. In the image shift algorithm (Algorithm 3), the individual facets were identified by the centroids of their reflections in each image. The inputs to the algorithm are the elevation and azimuth angle, the set of images to be assessed, the image channel, and the region of interest. The output of the algorithm is a set of centroids and a final image that contains all the correlated images taken during the scanning procedure.

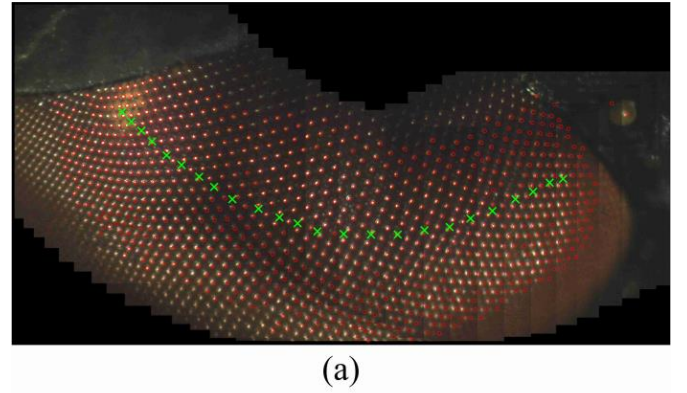

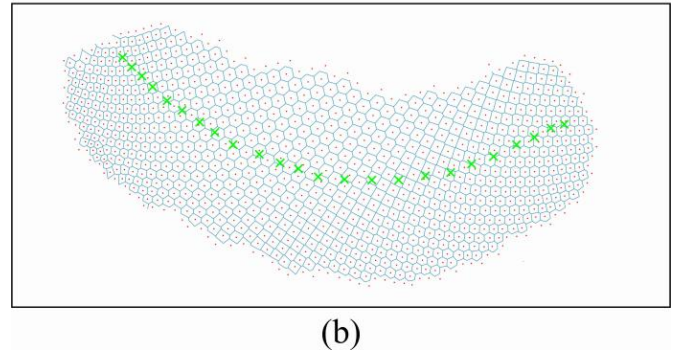

Fig. 7. The right eye of a fly scanned from one side to the other side. (a) Combined, overlapping images of an image series in which the azimuth was changed stepwise by 5°, together with the image centers (green crosses) and the facet centroids (red circles). (b) Voronoi diagram of the facet centroids with the image centers as in (a).

*F. Filling in missing facets*

Not all facets are identified by the centroid procedure, due to low reflectances caused by minor surface irregularities and specks of dust, and the latter can also cause erroneous centroids (Fig. 6a). This problem is resolved by applying the Filling-in algorithm (Algorithm 4). The centroids in an area are first determined (Fig. 6a), and then the FFT is calculated (Fig. 6b). The first ring of harmonics (yellow stars in Fig. 6b) defines three directions, indicated by the blue, red and green lines (Fig. 6b). Inverse transformation of the harmonics along the three directions yields the grey bands in Fig. 6c-e. Fitting a second order polynomial to the grey bands yields lines connecting the facet centroids along the three lattice axes, and the crossing points of the lattice lines then correspond to the facet centres.

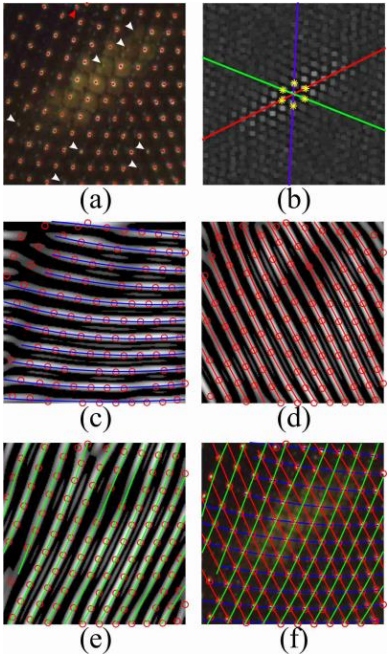

Fig. 6. Deriving missing facet centroids by applying Fourier transforms. (a) A local RGB-image with facet centroids (red dots). White arrowheads indicate missing facets, and the red arrowhead points to an erroneous centroid. (b) FFT of the centroids of (a) with the first ring of harmonics marked by yellow stars. (c-e) Inverse FFT of the centroids along the three directions indicated by the colored lines in (b), yielding the grey bands. The blue (c), red (d) and green (e) lines are quadratic polynomial fits to the grey bands, and the centroids (red circles) are those which were previously obtained prior to the Fourier transforms. (f) The fitted lines of (c-e) combined, together with the centroids of (a). The missing facet centroids are then derived from the crossing points.

The procedure is robust, because the example of Fig. 6 is an excessive case, as in most areas missing facets and erroneous centroids were rare.

*G. Scanning a fly eye*

We scanned a band of ommatidia across the eye by performing a series of stepwise azimuthal changes with $\Delta\varphi = 5°$, while holding the elevation $\theta$ at 30º. Scanning from the frontal side of the eye (Fig. 7a, right) to the lateral side (Fig. 7a, left) occurred in 24 steps. The centroids of the largely overlapping facet patterns were subsequently rotated by $\beta = \cos\theta\,\Delta\varphi$ and then, after shifting the centroids of each image with the vector obtained through algorithm 3 and filling in the missing facets with algorithm 4, the colocalized centroids were averaged. Figure 7a shows the combined images together with the image centers and facet centroids. Figure 7b shows the assembly of facets as a Voronoi diagram.

### III. CONCLUSION

We outlined an automated procedure for scanning insect eyes, which facilitates the assessment of the spatial organization of the visual axes and the ommatidial facet pattern. The case of a fly eye served as an example. Previous studies were performed by manually analysing images, using the pseudopupils as markers of the visual axes, while applying small bits of chalk dust to the eye as landmarks for determining the combined rotations and translations between successive images. In the present study, the addition of landmarks was unnecessary, as image rotations were known, and image correlation appeared to be sufficient for determining the translations. The filling-in algorithm was used to find facets that were not identified with the centroid procedure. The procedure at the same time identifies the facet lattice, which is essential in the further analysis of the lattice of visual axes. Here we have presented a partial scan of the eye. A more complete sampling and analysis of eye maps will be published elsewhere.

The applied key strategies relied on generalized methods such as the Fast Fourier Transformation, Gaussian filtering and binarization of captured images. Identification of facet centroids and pseudopupils enabled the automation of the scanning process. For instance, the binarization algorithm provides a set of properties such as regions, areas, and center of mass of captured images that are used as main guidance for the eye's automatic scanning.

Vision and navigation systems will benefit from the insights gained from the visual systems of insects that have been optimized by evolution for specific detection tasks. The aim of the present project is to enable the biomimetic or bioprincipic design of artificial compound eye systems. Exploring the physical properties and morphology of compound eyes may have substantial potential for medical and industrial vision applications.

### IV. APPENDIX

ALGORITHM 1
AUTOFOCUSING VIA FFT

$f_r$: **set** filter size for frequency domain of image
$R$: **set** region of interest of image
$C$: **set** desired image channel to blue
$Z$: **set** number steps for Z-axis motor
**for**: every step of the motor
    $I_R$: **get** image from region of interest $R$
    $I_C$: **extract** desired channel $C$ from image $I_R$
    $I_{FFT}$: **compute** FFT of image $I_C$
    $I_{shift}$: **shift** zero frequencies of $I_{FFT}$ to image plane center
    $I_{norm}$: **normalize** frequency domain of $I_{shift}$
    $C_I$: **find** center of frequency domain
    $f_h$: **apply** highpass filter to $I_{norm}$ around $C_I$
    $S_h$: **sum** values of filtered frequency domain
    **move** Z-axis motor over predefined number of steps
**end for**
$Z_h$: **find** Z-value where $S_h$ is maximal
**move** Z-axis motor to $Z_h$
$I_F$: **store** image with facet lens reflections obtained at $Z_h$

ALGORITHM 2
AUTOCENTERING VIA GAUSSIAN FILTERING AND BINARIZATION

$R$: **set** region of interest of image
$C$: **set** desired image channel to red
$Q$: **get** coordinate values of center of image plane
$P$: **get** center of pseudopupil
**while** (pseudopupil center is outside image center)
    $I_R$: **get** image from region of interest $R$
    $I_C$: **extract** desired channel from image $I_R$
    $F_G$: **apply** Gaussian filter to $I_C$
    $I_d$: **convert** filtered image to double (for binarization)
    $I_b$: **binarize** image $I_d$
    $P_b$: **extract** properties coordinates and area of binarized image $I_b$
    $P_p$: **compute** current position of pseudopupil represented by main object of $P_b$
    $E_{PQ}$: compute distance $PQ$
    $E_{XY}$: **convert** $E_{PQ}$ from pixels to $(X,Y)$ motor steps
    **if** $E_{XY} \neq 0$ then
        **move** X,Y-motors $E_{XY}$ steps
    **end if**
**end while**
$I_s$: **store** centered image

ALGORITHM 3
ASSESSING THE SHIFT OF FACET PATTERN IN SUBSEQUENT IMAGES

$N$: **define** the image set of $N$ images to be correlated
$\theta$: **define** the set of values of elevation steps
$\varphi$: **define** the set of values of azimuth steps
$\beta$: **compute** the set of image axes rotations
$R$: **set** region of interest of image
$C$: **set** desired image channel to blue
**for** every stored image ($i = \{1,\ldots,N\}$)
    $I_{S,i}$: **get** image from stored set of images
    $I_{rot,i}$: rotate image $I_{s,i}$ by $\beta$ around center
    $R_i$: **set** new region of interest of rotated image based on $I_{rot,i}$
    $I_{C,i}$: **select** blue channel of new region of interest of $R_i$
    $F_i$: **compute** and store facet centroids of $I_{C,i}$
    **for** $i = \{2,\ldots,N\}$,
        $D_i$: **compute** correlogram of images $I_{C,i-1}$ and $I_{C,i}$
    **end for**
    $I_{D,1-N}$: **merge** sequentially correlated images
**end for**

ALGORITHM 4
FILLING IN MISSING FACETS

$N$: **define** the images to be correlated
$C$: **set** desired image channel to blue
**for** every stored image ($i = \{1,\ldots,N\}$)
    $I_{s,i}$: **get** image from stored set of images
    $I_{C,i}$: **extract** desired channel from image $I_{s,i}$
    $I_{FFT}$: **compute** FFT of image $I_{C,i}$
    $(x_{r,i}, y_{r,i})$: **compute** positions of first ring of harmonics
    $a_{r,i}$: **compute** the three angular directions of positions $(x_{r,i}, y_{r,i})$
    $I_{H,i}$: **extract** harmonics along each of the three angular directions
    $I_{IFFT,i}$: **compute** IFFT of the three sets of harmonics $I_{H,i}$
    $C_{IF,i}$: **compute** and store properties of bands (regions) of $I_{IFFT,i}$
    $D_{IF,i}$: **fit** second order polynomial to IFFT bands of $C_{IF,i}$
    $I_{DC,i}$: **merge** fits and calculate crossing points
**end for**